# LIOUVILLE EQUATION IN STATISTICAL MECHANICS IS NOT APPLICABLE TO GASES COMPOSED OF COLLIDING MOLECULES

Huai-Yu Wang[a]

Department of Physics, Tsinghua University, Beijing 100084 China

**Abstract:** Liouville equation is a fundamental one in statistical mechanics. It is rooted in ensemble theory. By ensemble theory, the variation of the system's microscopic state is indicated by the moving of the phase point, and the moving trajectory is believed continuous. Thus, the ensemble density is thought to be a smooth function, and it observes continuity equation. When the Hamiltonian canonical equations of the molecules are applied to the continuity equation, Liouville equation can be obtained. We carefully analyze a gas composed of a great number of molecules colliding with each other. The defects in deriving Liouville equation are found. Due to collision, molecules' momenta changes discontinuously, so that the trajectories of the phase points are actually not continuous. In statistical mechanics, infinitesimals in physics and in mathematics should be distinguished. In continuity equation that the ensemble density satisfies, the derivatives with respect to space and time should be physical infinitesimals, while in Hamiltonian canonical equations that every molecule follows, the derivatives take infinitesimals in mathematics. In the course of deriving Liouville equation, the infinitesimals in physics are unknowingly replaced by those in mathematics. The conclusion is that Liouville equation is not applicable to gases.

[a] wanghuaiyu@mail.tsinghua.edu.cn

**Résumé :** L'équation de Liouville est fondamentale en mécanique statistique. Elle est ancrée dans la théorie des ensembles. Selon la théorie des ensembles, la variation de l'état microscopique du système est indiquée par le déplacement du point de phase, et la trajectoire de déplacement est considérée comme continue. Ainsi, la densité d'ensemble est considérée comme une fonction lisse, et elle observe l'équation de continuité. Lorsque les équations canoniques hamiltoniennes des molécules sont appliquées à l'équation de continuité, l'équation de Liouville peut être obtenue. Nous analysons avec soin un gaz composé d'un grand nombre de molécules entrant en collision les unes avec les autres. Les défauts de dérivation de l'équation de Liouville sont trouvés. En raison de la collision, la quantité de mouvement des molécules change de manière discontinue, de sorte que les trajectoires des points de phase ne sont en fait pas continues. En mécanique statistique, il faut distinguer les infinitésimaux en physique et en mathématiques. Dans l'équation de continuité que satisfait la densité d'ensemble, les dérivées par rapport à l'espace et au temps devraient être des infinitésimales physiques, tandis que dans les équations canoniques hamiltoniennes que chaque molécule suit, les dérivées prennent des infinitésimales en mathématiques. Au cours de la dérivation de l'équation de Liouville, les infinitésimaux en physique sont remplacés sans le savoir par ceux en mathématiques. La conclusion est que l'équation de Liouville n'est pas applicable aux gaz.

# I. INTRODUCTION

We deal with a gas composed of *n* molecules, where *n* is a large number. Here it is emphasized that the molecules are real ones, such as $H_2$, $CO_2$ and so on. The molecules collide with each other frequently. Besides the collision, there are interactions between them, which sometimes may be weak enough so that can be neglected.

For the sake of convenience, the inner degrees of freedom of the molecules are not taken into account. The inclusion of the inner degrees of freedom does not affect the final conclusion. The space coordinates and momentum of each molecule are respectively denoted as *q* and *p*.

Every molecule obeys the equations of motion of classical mechanics. The Hamiltonian of the system is denoted by *H*. The *i*-th molecule meets the following Hamiltonian canonical equations.

$$\dot{q}_i = \frac{\partial H}{\partial p_i}, \dot{p}_i = -\frac{\partial H}{\partial q_i}. \tag{1.1}$$

We briefly review the derivation process of Liouville equation. This is to facilitate us to point out below where in the process the confusions of concepts rise.

According to Gibbs' ensemble theory,[1–8] all the spatial and momentum coordinates of the *n* molecules constitute a 6*n*-dimensional phase space,[2–5] and a microscopic state of the system is represented by a phase point[5–7] in the phase space. Because all the molecules move, the microscopic state of the system varies with time. In the phase space, a phase point moves with time *t*, so that it is denoted as $(t, Q, P) = (t, q_1, \cdots, q_n, p_1, \cdots, p_n)$. The infinitesimal phase volume,[5,7] or volume element[9] in the phase space is defined by

$$d\Omega = dq_1 \cdots dq_n dp_1 \cdots dp_n. \tag{1.2}$$

Assume that there are *N* identical gases. That two systems are identical means that their Hamiltonians are the same.[2] Here, we distinguish the meanings of the *n* and *N*. The *n* is the number of molecules in a gas, and 6*n* are the number of axes that constitute the phase space. The *N* is the number of identical gases and is also the number of the phase points in the phase space. At an instant, the *N* gases are in different microscopic states. Therefore, the *N* phase points form a distribution in the phase space, and they move with time. The density of the phase point is denoted as $D(t, q_1, \cdots, q_n, p_1, \cdots, p_n)$, and its integration over the whole phase space is *N*, the number of the gases. The ensemble density is defined by[1,2,4–6]

$$\rho(t, q_1, \cdots, q_N, p_1, \cdots, p_N) = \frac{D(t, q_1, \cdots, q_N, p_1, \cdots, p_N)}{N}. \tag{1.3}$$

It is normalized in the whole phase space.

$$\int \rho dq_1 \cdots dq_N dp_1 \cdots dp_N = 1. \tag{1.4}$$

The moving trajectory of every phase point is thought to be continuous,[5] the trajectories of different points do not intersect when external field is absent. So, density current can be defined,

$$\boldsymbol{j} = \rho \boldsymbol{v}. \tag{1.5}$$

Naturally, the ensemble density is believed a smooth function in the phase space.[2] The points in the phase space can be regarded as a fluid. Consider a drop of the fluid. When the drop moves, it may change its shape, but its volume remains unchanged. This was expressed by[1]

$$\frac{d\rho}{dt} = 0. \tag{1.6}$$

It also means that the fluid is incompressible, so that the continuity equation of the ensemble density stands,

$$\frac{\partial \rho}{\partial t} + \nabla \cdot (\rho \boldsymbol{v}) = 0. \tag{1.7}$$

Here the gradient operator is

$$\nabla = (\frac{\partial}{\partial \boldsymbol{q}_1}, \cdots, \frac{\partial}{\partial \boldsymbol{q}_n}; \frac{\partial}{\partial \boldsymbol{p}_1}, \cdots, \frac{\partial}{\partial \boldsymbol{p}_n}). \tag{1.8}$$

Equation (1.7) can be explicitly written as

$$\frac{\partial \rho}{\partial t} + \sum_i (\frac{\partial}{\partial \boldsymbol{q}_i}(\rho \dot{\boldsymbol{q}}) + \frac{\partial}{\partial \boldsymbol{p}_i}(\rho \dot{\boldsymbol{p}})) = 0. \tag{1.9}$$

Applying (1.1) to (1.9) leads to

$$\frac{\partial \rho}{\partial t} + \sum_i (\frac{\partial \rho}{\partial \boldsymbol{q}_i}\dot{\boldsymbol{q}} + \frac{\partial \rho}{\partial \boldsymbol{p}_i}\dot{\boldsymbol{p}}) = 0. \tag{1.10}$$

This well-known Liouville equation.[1,4,5,10,11]

Liouville equation is a most fundamental one in statistical mechanics. It was used to prove theorems and derive formulas. For instance, in literature, starting from Liouville equation, Boltzmann transport equation can be derived,[3,4,6,8–17] which is called BBKGY method.

We think that the above derivation process should be inspected in detail. What is the real trajectory of a molecule frequently colliding with others has not been scrutinized yet. In section II, we show that the phase trajectories are not continuous. In section III, we distinguish physically and mathematically infinitesimals in calculation of physical quantities in the systems of statistical mechanics. These two kinds of infinitesimals are confused in deriving Liouville equation. Section IV is the further discussion of the utility of Liouville equation in the systems of statistical mechanics. Finally, section V is the conclusion. In Appendix, we provide reasons that time scale should not be as small as at will.

## II. PHASE TRAJECTORIES ARE NOT CONTINUOUS

Here we merely consider elastic collision between molecules. At the moment of collision, the total momentum and energy are conserved. The momenta of two colliding molecules before the collision is denoted by $\boldsymbol{p}_{10}$ and $\boldsymbol{p}_{20}$, and those after collision by $\boldsymbol{p}_1$ and $\boldsymbol{p}_2$. Then the conservations of the total energy and momentum are expressed by

$$\frac{1}{2m}\boldsymbol{p}_{10}^2 + \frac{1}{2m}\boldsymbol{p}_{20}^2 = \frac{1}{2m}\boldsymbol{p}_1^2 + \frac{1}{2m}\boldsymbol{p}_2^2 \tag{2.1a}$$

and

$$\boldsymbol{p}_{10} + \boldsymbol{p}_{20} = \boldsymbol{p}_1 + \boldsymbol{p}_2. \tag{2.1b}$$

In three-dimensional space, (2.1) contains four equations, but there are six quantities, $(p_{1x}, p_{1y}, p_{1z}), (p_{2x}, p_{2y}, p_{2z})$, to be solved. So, among the momenta of the two outgoing molecules, there are two uncertain quantities. For example, the angles $(\theta, \varphi)$ between the $\boldsymbol{p}_1$ and $\boldsymbol{p}_2$ can be random. Anyhow, a molecule's momentum has a sudden change during a collision.

A molecule's three coordinates and three momentum component axes constitute its generic or specific phase space, also called $\mu$–space.[6] Figures 1(a) and 1(b) respectively depict the spatial and momentum subspaces of the $\mu$–space of the $i$-th molecule. Let us see the trajectories of the molecule in the subspaces.

Suppose that there is no interaction between molecules other than collisions. At initial time, the molecule is at position A in the coordinate subspace and is doing uniform linear motion with momentum $\boldsymbol{p}_A$. Its trajectory in coordinate subspace is a piece of straight line, and that in momentum subspace is just a point located at $\boldsymbol{p}_A$. When it moves to position B, it collides with another molecule. At this instant, its location remains unchanged but its momentum suddenly changes to be $\boldsymbol{p}_B$. After that, it moves with $\boldsymbol{p}_B$ to another point C, and at this point collides with another molecule. Due to the collision, its momentum suddenly changes to be $\boldsymbol{p}_C$. Then it moves with $\boldsymbol{p}_C$ until next collision happens, and so forth. It is seen that in the spatial subspace, the trajectory of the molecule is pieces of straight lines, and the end of one piece is the starting point of the next piece, see FIG. 1(a). The length of each piece is the distance it goes between adjacent collisions. In the momentum subspace, the trajectories are discrete points, see FIG. 1(b).

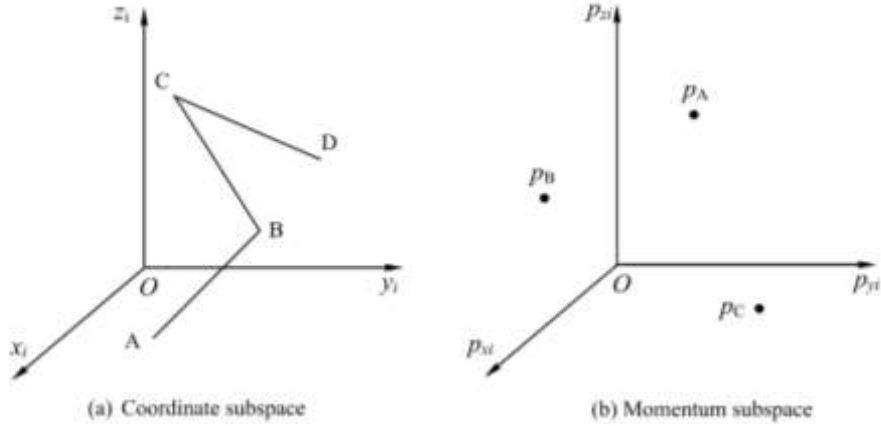

**FIG. 1.** The trajectory of the $i$-th molecule in the spatial and momentum subspaces of its $\mu$–space.

Here it is stressed that we are talking about real molecules but not hard spheres. For real molecules, the collision rules are just Eq. (2.1), and there is no other detailed information about how the two molecules contact during the collision. For a hard-sphere system, the dynamics is completely deterministic because the after-collision velocities are uniquely determined by the collision rule.[18,19] Even in this case, the momenta of the molecules before and after collisions are discontinuous.

If a molecule interacts with others in addition to the collisions, its momentum will vary when it moves. As a result, its trajectory between adjacent collisions in the spatial subspace becomes a piece of curve, and correspondingly, the point in the momentum subspace extends to be a piece of curve.

Anyhow, in the momentum sunspace, the trajectories before and after a collision are disconnected.

The same consideration applies to any other molecule's trajectory in the gas.

By the way, Eq. (6.12) in [4] and (2.74) in [11] are merely for the cases where there is no collision taking place.

Now, a phase point has $6n$ coordinates, among which $3n$ are spatial and $3n$ are momentum. When the phase point moves, its momentum components vary discontinuously, with sudden changes from time to time. Therefore, its phase trajectory is not continuous.

Let us consider a ball of phase points. When the ball moves, not only its shape changes, but also, some phase points jump out of and some others jump into the ball. Due to this reason, one is actually unable to define a density current by Eq. (1.5), neither can he regard the phase points as an incompressible fluid, i.e, Eq. (1.6) is invalid. Consequently, the continuity equation (1.7) does not hold.

In the gradient operator (1.8), the derivatives with respect to momenta require that momenta vary continuously. Since now the momenta of the molecules suddenly change frequently due to collisions, the operator (1.8) cannot be applied to (1.7).

Due to the collisions, the phase space is not a simply connected domain. In deriving Liouville equation, it is assumed that phase points' trajectories are continuous, which does not take the collisions into account. It was pointed out that Liouville equation itself does not have any collision effects.[16]

A function with the coordinates of the phase space as its arguments,

$$f(t, \bm{q}_1, \cdots, \bm{q}_N, \bm{p}_1, \cdots, \bm{p}_N), \tag{2.2}$$

is called a phase function.[4] For examples, the ensemble density (1.3) and probability current defined by (1.5) are phase functions. For a gas, its phase functions are not smooth ones.

## 3. DISTINGUISHING PHYSICAL AND MATHEMATICAL INFINITESIMALS

### A. Infinitesimal in mathematics

In evaluation of physical quantities, derivatives are often employed. In mathematics, the concept of derivative is as follows. Suppose that a quantity $y$ is a function of argument $x$, $y(x)$. The derivative of $y$ with respect to $x$ is defined by

$$\frac{dy}{dx} = \lim_{\Delta x \to 0} \frac{y(x+\Delta x) - y(x)}{\Delta x} = \lim_{\Delta x \to 0} \frac{\Delta y}{\Delta x}. \tag{3.1}$$

Here the argument $x$ must be a continuous one. That $\Delta x \to 0$ requires that $\Delta x$ approaches zero continuously without sudden change, and $dx$ is infinitely small. Consequently, $\Delta y \to 0$, and $y$ also approaches zero continuously. Both the numerator and denominator in (3.1) are called infinitesimals in mathematics.

In Hamiltonian canonical equations (1.1) that a molecule's motion obeys, the derivatives take infinitesimals in mathematics.

### B. Macroscopic and microscopic motions

In nature, there are macroscopic and microscopic motions.

Microscopic motion means those of single molecules. Usually, microscopic motion is described by microscopic physical quantities, such as a molecule's momentum, energy, and so on, which are also called micro-physical quantities in short. For instances, the momenta of molecules, such as those in Eq. (2.1), and a molecule's velocity $\bm{v} = \dfrac{d\bm{r}}{dt}$ are micro-physical quantities, where the derivatives should take mathematical infinitesimals, also called microscopic infinitesimals, or in short, micro-infinitesimals.

A large number of molecules constitute a macroscopic body the motion of which is termed as macroscopic motion. Macroscopic motion is described by macroscopic physical quantities, such as density, pressure, temperature, internal energy, and so on, which are also called macro-physical quantity in short. Some macro-physical quantities can be measured by devices.

The system in investigation in this work is constituted by molecules which distribute in space discretely. Such a microscopic discreteness imposes constraints on the evaluation of macro-physical quantities.

### C. Macroscopic infinitesimal in space

The evaluation of macro-physical quantities involves derivatives either. For instance, molecule number density, i.e., molecules' number in unit volume is defined by

$$\frac{\mathrm{d}n}{\mathrm{d}V} = \lim_{\Delta V \to 0} \frac{\Delta n}{\Delta V}. \tag{3.2}$$

In mathematics, both numerator and denominator in (3.2) should be infinitely small. Nevertheless, there are rooms between neighboring molecules. When the volume $\Delta V$ goes continuously to mathematical infinitesimal, the molecule's number $\Delta n$ in this volume does not decrease continuously, and the least nonzero number is 1. Therefore, it is impossible to take $\Delta V$ as a mathematical infinitesimal. One can merely take it macroscopically small enough, but microscopically still a finitely large volume inside which there are sufficiently many molecules. Thus, the infinitely small volume in (3.2) is not really infinitesimal in mathematics. It is called macroscopic infinitesimal, in short macro-infinitesimal, also called infinitesimal in physics. Correspondingly, the $\Delta n$ in numerator is also a macro-infinitesimal which is a finite value in microscopic view. In a word, the size of the volume element in (3.2) should be much larger than the distance between neighboring molecules.

In the subsection 2.13 in titled "Distinguishing the physics from the mathematics" in [20], the distinguish between infinitesimals in mathematics and in physics was emphasized with electric charge density being an example.

Let us see another example. A function $f(\boldsymbol{q},\boldsymbol{p},t)$ is defined as the distribution function of molecules, such that the molecule number in $\mathrm{d}\boldsymbol{q}\mathrm{d}\boldsymbol{p}$ is

$$f(\boldsymbol{q},\boldsymbol{p},t)\mathrm{d}\boldsymbol{q}\mathrm{d}\boldsymbol{p} = f(\boldsymbol{q},\boldsymbol{p},t)\mathrm{d}V\mathrm{d}\boldsymbol{p}, \tag{3.3}$$

where $\mathrm{d}\boldsymbol{q} = \mathrm{d}V$ is the spatial volume element. The integration of $f(\boldsymbol{q},\boldsymbol{p},t)$ over the momentum space results in the molecule number density $\rho(\boldsymbol{q},t)$ in real space.[21] It was mentioned[10] that the volume elements $\mathrm{d}V$ and $\mathrm{d}\boldsymbol{p}$ are not to be taken literally as mathematically infinitesimal quantities. They are finite volume elements which are large enough to contain a very large number of molecules and yet small enough so that compared to macroscopic dimensions they are essentially points. That such a choice is possible can be seen by an example. Under standard conditions there are about $3\times10^{19}$ molecules/cm$^3$ in a gas. If we choose $dV \sim 10^{-10}$ cm$^3$, which to us is small enough to be called a point, there are still the order of molecules $3\times10^9$ in $dV$.

An infinitesimal volume element $dV$ really means one that is not mathematically but physically small, i.e., a region of space which is very small in comparison with the characteristic dimension $L$ of the problem, but still large in comparison with molecular dimension.[9]

Because in (3.2) both the denominator and numerator are physically infinitely small, the calculated molecule density is a macroscopically continuous function.

If one has to fastidiously investigate the variation of the density in a sense of microscopic infinitesimal, then, strictly speaking, density, pressure and other macroscopic quantities are of discreteness. In this sense, entropy is also of discreteness,[22] because the evaluation of entropy involves counting molecule number. Similarly, pressure, enthalpy, free energy and so on are all macroscopically continuous but microscopically discrete. The feature of a thermodynamic system is that it is constituted by a large number of molecules, so that one can consider its macro-physical quantities as continuous ones, disregarding their microscopic discreteness.

We return to the phase space. In the phase point fluid, the points are discrete units and there are rooms between them. When a volume element in the phase space, Eq. (1.2), shrinks continuously, the variation of the phase point number inside this element is not continuous. This volume element thus ought to be macroscopically infinitely small, just as that in (3.2).

Therefrom, in the gradient operator (1.8), the derivatives to spatial coordinates take macroscopic but not microscopic infinitesimals. A volume element still contains many molecules. However, in writing Liouville equation (1.10), Hamiltonian canonical equation (1.1) are employed, where the derivatives with respect to spatial coordinates take microscopic infinitesimals. It is seen that from (1.7) to (1.10), spatial macroscopic infinitesimals are unknowingly replaced by microscopic ones. Thus, the derivation process is incorrect.

When one discusses the motion of molecules, only microscopic infinitesimals are involved. When he investigates the motion of a system constituted by a large number of molecules, macroscopic infinitesimals are concerned. The examples are thermodynamics, fluid mechanics, elastic mechanics, continuum medium mechanics, aerodynamics and so on.

In each of microscopic and macroscopic fields, there will not occur the mix of macroscopic and microscopic infinitesimals.

Statistical mechanics is a special discipline where objects under research are macroscopic systems composed by a large number of molecules and where microscopic and macroscopic motion have to be dealt with simultaneously. There can be a lot of microscopic states corresponding to a macroscopic state. Hence, in statistical mechanics, people consider both macroscopic and microscopic states. As a comparison, in thermodynamics, only macroscopic states are considered.

To describe macroscopic (microscopic) states, macro(micro)-physical quantities are needed. That is to say, in this discipline both macro- and micro-physical quantities are to be calculated. Roughly speaking, a macro-physical quantity is in some way the statistical average of micro-physical quantities. For instance, the total kinetic energy is the statistical average of the kinetic energies of all the molecules in the system. Although there may be certain relationships between macro- and micro-physical quantities, they should not replace each other directly.

The derivatives with respect to coordinates involved in calculating macro (micro)-physical quantities takes macroscopic (microscopic) infinitesimal in space. In deriving Liouville equation, the macroscopic infinitesimals in Eq. (1.9) are replaced by microscopic ones in (1.1). This is an example of the confusion of macroscopic and microscopic infinitesimals.

It is actually well known that spatial macroscopic and microscopic infinitesimals should be distinguished, but people may not pay a particular attention in calculation, as exposed in the derivation process of Liouville equation. As a matter of fact, temporal macroscopic and microscopic infinitesimals should also be distinguished. This has been seldom mentioned.[3,23] In the next subsection we give the reason.

**D. Macroscopic infinitesimal in time**

Measuring macro-physical quantities has to resort to devices which are also macroscopic systems, such as thermometer, piezometer, and so on. Measurement is implemented through the interactions, usually collisions, between molecules in the measured system and device.

Measuring a macro-physical quantity, no matter what device is used, needs a period of time but not an instant. Here an instant means an infinitely small piece of time, a mathematical infinitesimal.

For instance, measuring a system's temperature needs a period of time, within which the molecules in the system fully collide with those in the device so as to have adequate exchange of heat between the system and device. The measuring time can be very short, say 0.1, or even 0.01 seconds, which is macroscopically short enough. Nevertheless, from the microscopic point of view, this time is long enough, since within the 0.01 seconds, a molecule can collide with others by about $10^7$ times. This measurement time, 0.01 seconds, can be regarded as an instant, which is called macroscopic instant.

A measurement consists in recording a time average of the property in question in the sample of the ensemble, with specified dynamical state at time $t$, over an interval of time $T$, macroscopically short, but microscopically long in a sense presently to be made more precise.[3]

"Bogoliubov ….. defines three time intervals, which we label $\tau_1$, $\tau_2$ and $\tau_3$. In the interval $\tau_1$, two molecules are in each other's interaction domain. The interval $\tau_2$ is the mean-free-collision times, which is the mean times between collisions. The time $\tau_3$ is the average time taken for a molecule to traverse the container in which the gas is confined. For a mole sample a gas confined to a macroscopic container, we may write $\tau_1 \ll \tau_2 \ll \tau_3$ ."[8] Roughly speaking, $\tau_1$ and $\tau_3$ correspond to micro- and macro-infinitesimal times, respectively.

When describing the variation of a macro-physical variable, the spatial interval has to be taken as macro-infinitely small. Correspondingly, time interval has to be taken as macro-infinitely small either.

Therefore, in Eq. (1.7), the gradient operator in the second term should take macroscopic infinitesimals. Correspondingly, in the first term, the derivative with respect to time should do either. However, in the first term in Eq. (1.10), the derivative to time unknowingly takes microscopic infinitesimal.

In one word, in deriving Liouville equation, both spatial and temporal macroscopic and microscopic infinitesimals are confused.

The conclusion is that Liouville equation does not apply to gases.

One may argue that a time scale that is even finer than the micro one can help to present a more detailed description of collisions such that the changes of momenta are continuous. In Appendix, we address three points.

## IV. DISCUSSION

By Fig. 1, we have been clear that when a phase point moves, its momenta can change suddenly due to the collisions between molecules. Hence, the ensemble density is not a smooth function for its arguments have sudden changes.

There are three terms in Liouville equation (1.10). Due to collisions, molecules' momenta have sudden changes, so that the third term is incorrect. In the phase space, the phase volume element should be taken as macro-infinitely small, and so the second term is not right. Correspondingly, the derivative to time in the first term should take macroscopic infinitesimal.

In the course of deriving Liouville equation from continuity equation, the spatial and temporal macroscopic infinitesimals were unknowingly replaced by microscopic ones. This deriving process was illegal. For a gas, one is even unable to define the current density in phase space by Eq. (1.5) so that Eq. (1.6) is not guaranteed.

In the literature, Liouville equation is usually employed to carry out some theoretical analysis. A premise of deriving Liouville equation is that the trajectories of the phase points are assumed continuous, but this is not true for gases. Therefore, all the discussion based on continuous trajectories or Liouville equation are not valid.

The current defined by Eq. (1.5) in the phase space is also called probability current. When discussing the variation of the microscopic state of a system, the probability current is classified. The simplest ones are ergodic flow and mix flow.[4] They are related to the discussion of ergodic hypothesis. Birkhoff[24] established a criterion for determining if a system is ergodic, called ergodic theorem. In order to arrive at this criterion, the ensemble density was assumed smooth, although Liouville equation was not mentioned. We have shown above that the ensemble density is not a smooth function.

The proof of the famous Poincare recurrence theorem[8] is based on the continuity of the probability current in the phase space. Since for a gas, the trajectories of the phase points are not continuous, this proving process of Poincare recurrence theorem is not right.

People have tried to give a theoretically satisfactory derivation of Boltzmann equation, and thought that Liouville equation was a right starting point. The derivation process is called BBKGY method. Now that for gases Liouville equation is not correct, the deriving process by the BBKGY method is not valid.

There are two points in the BBKGY procedure showing serious problems. One is that as has been mentioned above: Liouville equation itself does not embody collision[16] while Boltzmann equation has a collision term. The other is that similar to Newton's equation of motion, $F = \frac{dp}{dt}$, Liouville equation (1.10) is a differential equation so that is time-reversible, while Boltzmann equation reflects time-irreversibility unless the system is in equilibrium states. Therefore, the derivation of Boltzmann equation from Liouville equation means that a collision term is created from a collisionless equation, and time-irreversibility is created from a time-reversible equation.

By the way, here we like to distinguish two different distribution functions in the BBKGY deriving process and even in Boltzmann equation itself.

To do so, we first put down Boltzmann transport equation. In Eq. (3.3), a distribution function $f(\boldsymbol{q},\boldsymbol{p},t)$ was defined which represented the mean number of molecules in unit phase space $d\boldsymbol{q}d\boldsymbol{p}$.[4,8,9] Boltzmann equation is[4,6,8,9]

$$\frac{\partial f}{\partial t} + \boldsymbol{v}\cdot\nabla f = \int (ff_1' - ff_1)w'(\boldsymbol{p},\boldsymbol{p}_1;\boldsymbol{p}',\boldsymbol{p}_1')d\boldsymbol{p}_1 d\boldsymbol{p}_1' d\boldsymbol{p}', \qquad (4.1)$$

where $w'$ is defined through the cross section of two-molecule collision

$$d\sigma = \frac{w'(\boldsymbol{p},\boldsymbol{p}_1;\boldsymbol{p}',\boldsymbol{p}_1')}{|\boldsymbol{v}-\boldsymbol{v}_1|} d\boldsymbol{p}_1' d\boldsymbol{p}'. \qquad (4.2)$$

For the sake of simplicity, in Eq. (4.1) external fields are not taken into account. The distribution function on the left hand side of (4.1) is what defined in (3.3), and its integration over the $d\boldsymbol{q}d\boldsymbol{p}$ space is the total number of the molecules in the system.

$$\int f(t,\boldsymbol{q},\boldsymbol{p})d\boldsymbol{q}d\boldsymbol{p} = n. \qquad (4.3)$$

Please note that $f(\boldsymbol{q},\boldsymbol{p},t)$ does not follow the tracks of any specific molecule, but merely enumerates the molecule number within a macro-infinitely small range of spatial and momentum coordinates at any time. The distribution function is a macroscopically continuous one, and subsequently, the derivatives in (4.1) should take macro-infinitely small. The distribution function $f(\boldsymbol{q},\boldsymbol{p},t)$ is believed to obey one-molecule Liouville equation.[8,9]

Another distribution function is for single molecules. Consider Fig. 1. At a moment, one molecule is at a position in its own μ-space. At the next moment, it is at another position. After a sufficient long time, these point positions form a distribution in the μ-space. This distribution still depends on time, and denoted by $f^{(1)}(\boldsymbol{q}_1,\boldsymbol{p}_1,t)$ where the subscript 1 of the arguments refers to the first molecule. This function represents the probability density of the first molecule at the position $(\boldsymbol{q}_1,\boldsymbol{p}_1)$ in its μ-space at time $t$, no matter what positions of other molecules are. Hence, it is

$$f^{(1)}(\boldsymbol{q}_1,\boldsymbol{p}_1,t) = n\int \rho(\boldsymbol{q}_1,\boldsymbol{q}_2\cdots\boldsymbol{q}_N;\boldsymbol{p}_1,\boldsymbol{p}_2\cdots\boldsymbol{p}_N;t)\mathrm{d}\boldsymbol{q}_2\cdots\mathrm{d}\boldsymbol{q}_N\mathrm{d}\boldsymbol{p}_2\cdots\mathrm{d}\boldsymbol{p}_N. \qquad (4.4)$$

The function $f^{(1)}(\boldsymbol{q}_1,\boldsymbol{p}_1,t)$ defined by (4.4) is called generic distribution function,[6] or a one-body phase function.[4] The integration of $f^{(1)}(\boldsymbol{q}_1,\boldsymbol{p}_1,t)$ over momentum space does not have the meaning of molecule number density, but the distribution of the probability density of this molecule in real space. It is naturally believed that if all the molecules in a system are the same, then they have the same generic distribution, $f^{(1)}(\boldsymbol{q}_1,\boldsymbol{p}_1,t) = f^{(1)}(\boldsymbol{q}_2,\boldsymbol{p}_2,t)\cdots$.

We stress that the arguments of $f$ in (3.3) and $f^{(1)}$ in (4.4) have different meanings: the latter are those of the first molecule while the former do not belong to any specific molecule. In practice, it is postulated that the functions $f$ and $f^{(1)}$ are the same,

$$f(\boldsymbol{q},\boldsymbol{p},t) = f^{(1)}(\boldsymbol{q}_i,\boldsymbol{p}_i,t), i = 1,2,\cdots,n, \qquad (4.5)$$

although no one has explicitly pointed out this. In the BBKGY hierarchy the left hand side of (4.1) is $f^{(1)}(\boldsymbol{q}_1,\boldsymbol{p}_1,t)$.[3,4,6,8–17] Therefore, the postulation (4.5) is implied.

As a matter of fact, when Boltzmann proposed the equation (4.1), the postulation (4.5) had been implied already. In order to put down the collision integral, in the right hand side of (4.1), two-molecule collision number, $N(\boldsymbol{p}\boldsymbol{p}_1 \to \boldsymbol{p}'\boldsymbol{p}'_1)$, is needed. It is assumed that

$$N(\boldsymbol{p}_1\boldsymbol{p}_2 \to \boldsymbol{p}'_1\boldsymbol{p}'_2) \propto f^{(1)}(\boldsymbol{q}_1,\boldsymbol{p}_1,t)f^{(1)}(\boldsymbol{q}_2,\boldsymbol{p}_2,t). \qquad (4.6)$$

That is to say, it is proportional to the generic distribution functions of the two molecules before collision. Therefore, the distribution function $f$ in the left hand side of (4.1) is that of the left hand side of (4.5), while the functions $f$ in the integral of the right hand side of (4.1) are just that of the right hand side of (4.5). Therefore, the postulation (4.5) has been implied.

In the above discussion, the inner degrees of freedom of the molecules are not taken into account. If they are, the conclusions above are retained. For example, if every molecule has an angular momentum, there will be the equation of the total angular momentum conservation during the collision in addition to Eq. (2.1). There are three such equations for three Cartesian coordinates but six angular momentum components of the two outgoing molecules to be resolved. As a result, there are three more undetermined quantities. Still, the changes of the molecules' momenta before and after collision are discontinuous.

At last, two points have to be pointed out.

The first point is that although the reasoning or derivation processes using Liouville equation is not valid. The conclusions that the reasoning process wants to prove may still be correct.

For examples, the ergodic hypothesis may be correct, but it cannot be proved by means of the premise that ensemble density is a smooth function for a gas. The ergodic hypothesis may be correct for a gas. The reason may be that after a collision, a molecule's momentum stochastically distributes, as illustrated by Fig. 1.

For a collision-free system, interactions between molecules are always finite, so that the ergodic hypothesis may not be established. An example is a coupled oscillator system,[2] the lack of ergodicity in which was proved rigorously.[25] Numerical computation showed that for some energy values, a nonharmonic system is not ergodic,[26] where the number of coupled oscillators was few and there was no collision between the oscillators. In this case, the spatial and momentum coordinates of each oscillator vary continuously, and the derivatives of spatial and momentum coordinates of each oscillator in (1.10) can take mathematic infinitesimals. Consequently, Liouville equation stands for such a system.

Poincare recurrence theorem may be correct, but the proof process based on the continuity of the phase trajectories is illegal.

The derivation of Boltzmann equation does not necessarily need Liouville equation.[6,8,9] Boltzmann himself established the equation from a phenomenological point of view. The BBKGY method has severe problems mentioned above.

The second point is that the invalidity of Liouville equation for gases does not substantially affect the evaluation of physical quantities of the systems.

In fact, the fundamental laws of thermodynamics and statistic mechanics were achieved without need of Liouville equation.

For examples, before the ensemble theory emerged, Maxwell velocity distribution function of gases, Boltzmann transport equation, and Boltzmann $H$ theorem had been derived. When the Boltzmann $H$ theorem was applied to an equilibrium state, Boltzmann distribution was derived, and in the case of quantum systems, Fermi-Dirac and Bose-Einstein distributions for identical particles could be obtained.[8,27] Moreover, in history, Boltzmann, Fermi-Dirac, and Bose-Einstein distributions were obtained respectively independently without use of ensemble theory.

In statistical mechanics textbooks, Liouville equation may be introduced, but after that, it does not have further use except deriving Boltzmann equation by BBKGY method. The application of Liouville equation is merely to embody the theoretical perfectness in discussing some topics based on the ensemble theory.

In this work, we consider classical molecules. In Eq. (2.1), if we choose the angles $(\theta, \varphi)$ between the momenta of the two molecules after the collision, the two outgoing momenta distribute randomly in spatial directions. In quantum mechanics, after scattering, a molecule's wave function can spread in all directions in space, because the scattered wave function is a spherical one at a distance.[28–30] Thus, the qualitative conclusion in quantum mechanics will be the same as that in classical mechanics. The detailed investigation of the case of quantum mechanics is to be undergone later.

## V. CONCLUSION

In this work, we carefully inspect the derivation process of Liouville equation. We investigate gases constituted by a large number of real molecules, with collisions between them. Every molecule obeys Hamiltonian canonical equations. Because of the collisions, the molecules' momenta can change discontinuously. Consequently, the trajectories of the phase points are not continuous, and phase functions are not smooth. For such gases, Liouville equation is not applicable, since the derivation process assumes that ensemble density is a smooth function in the phase space and observes continuity equation. Furthermore, the derivatives in the continuity equation should take spatial and temporal infinitesimals in physics, while those in Hamiltonian conical equations can take infinitesimals in mathematics. In the process of deriving Liouville equation, physical infinitesimals are unknowingly replaced by mathematical ones. The conclusion is that Liouville equation is not valid for gases.

For systems composed of few molecules without collisions between them, Liouville equation may be applicable, such as coupled oscillator systems.

In proving ergodic theorem and Poincare recurrence theorem, the prerequisite is that the ensemble density is a smooth function. Thus, the proving processes are illegal. In spite of the illegalness of the proving process, these two theorems themselves may be correct, although rigorous proofs of the theorem are still desired.

The BBKGY method, the derivation process of Boltzmann equation from Liouville equation, implies that a collision term is created from a collisionless equation, and time-irreversibility is created from a time-reversible equation. Hence, the derivation process is illegal. Nevertheless, Boltzmann equation itself is correct.

Fortunately, in evaluation of the physical quantities in thermodynamics and statistical mechanics, Liouville equation is not indispensable. Therefore, although Liouville equation is not applicable to gases, this does not substantially affect the investigation of the systems.

**Acknowledgements**

This work is supported by the National Key Research and Development Program of China [Grant No. 2018YFB0704304].

## APPENDIX: IT IS UNREASONABLE TO SET A FINER TIME SCALE IN ORDER TO INVESTIGATE THE PROBLEM OF COLLISION

When discussing collision between molecules, one may argue that it is possible that we set a sufficiently short time scale such that every detail during the collision will be known and the molecules' momenta will vary continuously. We address the following three points.

First, at the sufficiently short time scale, the motion of a gas would be strange.

Let us consider a gas under standard conditions. In this gas, the time interval between two adjacent collisions of a molecule with others is denoted as $\tau$, and the mean free path of a molecule denoted as $\lambda$. Then, $\tau \sim 10^{-11}$s and $\lambda \sim 10^{-7}$m.[10] The macro-infinitesimal time scale used to describe the motion of the gas is denoted as $\Delta t$. This scale should be much larger than $\tau$, i.e., $\Delta t \gg \tau$, say $\Delta t \sim 10^{-7}$s. Subsequently, the motion of the phase points in the phase space should also be described by this time scale. During the time period $\Delta t$, a molecule travels a distance $\sim 10^{-3}$m which is much greater than the $\lambda$. Under this time scale, the microstate of the gas changes rapidly, and the motion of the phase points in the phase space is believed like a fluid.

We have pointed out in the text that the momentum of every molecule changed rapidly due to frequent collisions. So, the movement of the phase points in the phase space should not be regarded as an incompressible fluid.

One may think that the reason that the sudden changes of the molecules' momenta is because the time scale chosen is too rough. If we set a much finer time scale, the situation will be different. Now, let us do so. The duration of the collision between two molecules is not zero, but a finite period of time, denoted as $\tau_1$, which is certainly much less than $\tau$: $\tau_1 \ll \tau$. We reasonably take $\tau_1 \sim 10^{-2}\tau$, i.e., $\tau_1 \sim 10^{-13}$s. In order to make all the collision details clear, we have to choose a time scale, denoted as $\Delta t_1$, which has to be much less than the collision time $\tau_1$: $\Delta t_1 \ll \tau_1$. For instance, we take $\Delta t_1 \sim 10^{-1}\tau_1$, i.e., $\Delta t_1 \sim 10^{-14}$s. Under such a scale, we assume that at every instant during the collision, the details of the interaction between the molecules are known, and the momentum of a molecule can be solved by means of Newtonian equation. In this way, we will see that the momentum varies continuously. Then, what will happen after the collision? Because the time scale is $\Delta t_1$, we have wait a time as long as $10^3 \Delta t_1$ to see next collision of this molecule. That is to say, the molecules move rather "slowly". The gas almost stagnates. Its microstate remains almost unchanged. Thus, it is seen that the time scale $\Delta t_1$ is improper to describe the motion of a gas and of other microscopic bodies, as well as of an ensemble. There would be no concepts related to macroscopic physical quantities such as temperature, pressure, and so on.

Please note that we must choose one of the scales $\Delta t$ and $\Delta t_1 \sim 10^{-3} \Delta t$, not both, to describe the motion of a gas. It is concluded that we have employ macro- (micro-) infinite small time scale for describing the motion of macro- (micro-) systems. When using the scale $\Delta t_1$, we are unable to describe the motion of a gas. When macro-infinitesimal time scale is employed to describe the motion of a gas, the momentum changes of the molecules due to collisions are abrupt.

Second, the Newtonian mechanics applies on the time scale that is macro-infinitely small.

We have assumed above that on a sufficiently small time scale, "during the collision, the details of the interaction between the molecules are known, and the momentum of a molecule can be solved by means of Newtonian equation." This assumption was wrong. The Newtonian equation applies merely on the macro-infinitely small time scale, but not on any very fine scale. When the time scale becomes finer, the collision will inevitably involve the details of the interactions between microscopic particles that constitute the molecules. At this level, quantum mechanics, instead of Newtonian mechanics, has to be resorted to.

Third, even in quantum mechanics, the collision makes molecules' momenta change suddenly.

Quantum mechanics can help to put down the wave functions of the colliding particles before and after the collision. Even so, we are unable to know what happened during the collision. Taking Compton scattering as an example, we can know, by measurements, the momenta and energies of the electron and photon before and after the collision, but we are unable to know the details how the energy and momentum transfer between the electron and photon during the collision. When the collision occurs, the momenta and energies of the electron and photon change suddenly, not continuously.

More radically, a pair of electron and positron collide and are annihilated. Meanwhile, a pair of photons are created. We are unable to tell the details that what happened during the collision: how the electron and positron disappear and how the photons appear. What we can know are the momenta and energies of them before and after the collision by measurement.

In short, collision is a very special interaction process. The details during the collision cannot be made clear by means of setting finer time scale. The collision makes the molecules' momenta change abruptly. Fortunately, there are conservation laws of momentum, energy, angular momentum, and so on. These laws help us obtain enough information to describe the motion of matters. The details of what happen during the collision are actually not need.